\documentstyle[sprocl,psfig,epsfig]{article}
\newcommand{\gc}{\gamma}
\newcommand{\gd}{\delta}
\newcommand{\gve}{\varepsilon}

\newcommand{\mbf}[1]{\mbox{\boldmath$#1$}}

\begin{document}
\renewcommand{\thefootnote}{\fnsymbol{footnote}}
\setcounter{footnote}{0}

\title{DEUTERON FORMATION IN HEAVY ION COLLISIONS \\WITHIN THE FADDEEV
  APPROACH}

\author{M. BEYER\footnote{Talk presented by M. Beyer at the workshop
    on ``Kadanoff-Baym equations'' -- progress and perspectives for
    many-body physics.}, C. KUHRTS, G. R\"OPKE}

\address{Fachbereich Physik, University of Rostock, 18051 Rostock,
  Germany}

\author{P.D. DANIELEWICZ}

\address{NSCL, Michigan State University, East Lansing, MI 48824, U.S.A.}

\maketitle
\begin{abstract}
  We address the formation of correlations in a general nonequilibrium
  situation. As an example we calculate the formation of light
  composite particles in a heavy ion collision. In particular we study
  the formation of deuterons via three-body reactions in some detail.
  To calculate the relevant reaction rates we solve the Faddeev
  equation which is consistently modified to include the self energy
  corrections and Pauli blocking factors due to  surrounding
  nucleons. We find that the  time scales of the reaction as well as
  the number of emitted deuterons change as medium dependent
  rates are used in the calculation.
\end{abstract}
\section{Introduction}
The description of the dynamics of an interacting many-body system is
particularly difficult when the quasiparticle approach reaches its
limits.  That may be the case when the residual interaction is strong
enough to build up correlations.  An example of such a system with
correlations is nuclear matter.  In heavy ion collisions, the nuclear
matter is first excited and compressed and then decompressed.  On a
macroscopic scale, the nuclear matter is formed during a supernova
collapse and becomes the material of which a neutron star is made.

Here, we address the formation of correlations in a general
nonequilibrium situation.  The simplest process in nuclear matter is
the formation of deuterons; their number is, besides the numbers of
other composite particles (tritons, helium-3's, $\alpha$-particles),
an important observable of heavy ion collisions. A microscopic
approach to treat this complicated process uses the Boltzmann
equation. In this context the equation has been numerically solved
utilizing the Beth-Uehling-Uhlenbeck approach, see e.g.
Ref.~\cite{dan91}. For nucleon ($f_N$) and deuteron ($f_d$) Wigner
distribution functions, the Boltzmann equation reads
%\clearpage
\begin{eqnarray}
\partial_t f_N+\{U,f_N\}&=&
{\cal K}^{\rm in}_N[f_N,f_d]\,(1-f_N)
-{\cal K}^{\rm out}_N[f_N,f_d]\, f_N,
\nonumber\\
\partial_t  f_d+\{U,f_d\}&=&
{\cal K}^{\rm in}_d[f_N,f_d]\,(1+f_d)
-{\cal K}^{\rm out}_d[f_N,f_d]\,f_d,
\label{eqn:Boltz}
\end{eqnarray}
where $U$ is a mean-field potential and $\{\cdot,\cdot\}$ denotes the
Poisson brackets. The collision integrals ${\cal K}$ appearing in
Eq.~(\ref{eqn:Boltz}) couple the distribution functions and respect
all the collisions (elastic and reactive) between the constituents. We
focus on the reaction part only, e.g. the loss reaction for the
deuteron is given by
\begin{eqnarray}
{\cal K}^{\rm out}_d(P,t)&=&
\int d^3k\int d^3k_1d^3k_2d^3k_3\;
|\langle k_1k_2k_3|U_0|kP\rangle|^2_{dN\rightarrow pnN}\nonumber\\&&
\qquad\times
\bar f_N(k_1,t)\bar f_N(k_2,t)\bar f_N(k_3,t)f_N(k,t)\nonumber\\&&
+\dots\label{eqn:react2}
\end{eqnarray}
where $\bar f= 1\pm f$ for bosons/fermions. The transition matrix
element is for the break-up operator $U_0$ and given by the solution
of the Alt-Grassberger-Sandhas (AGS) equation~\cite{alt67} explained
below.  Several approximations have been used to describe the
transition matrix $U_0$, e.g. Born approximation that may be justified
in the context of weak interactions, or impulse approximation that is
justified for higher scattering energies. One further strategy
is to introduce the cross section in the above equations
and use experimental values or meaningful extrapolations (assuming
detailed balance for the back reaction).

\section{In-medium three-body reactions}
As nuclear matter provides a dense system we address the question to
what extend the reaction {\em depends on the embedding medium}. To
this end we have derived an AGS-type equation for the three-particle
correlation embedded in an uncorrelated
medium~\cite{bey96,bey97,habil}. This will be shortly sketched here.
The equation is derived in the context of the cluster expansion or
Dyson equation approach~\cite{duk98}. The expansion is consistent in
the sense that the respective in-medium one- and two-body problems are
solved and implemented in the three-body equation, i.e. nucleon
self-energies $\gve$, Pauli blocking factors, and the Mott effect
of the deuteron are all consistently included. The c.m. momentum is
treated using four-body kinematics. Note, that the c.m. momentum
enters only parametrically into the three-body equation. The
three-body Green function evaluated in an uncorrelated medium is then
given by
\begin{eqnarray}
\nonumber
G_3(z)&=&\frac{\bar{f_1}\bar{f_2}\bar{f_3}+f_1f_2f_3}
{z- \gve_1-\gve_2-\gve_3}\\[1ex]
&&+\frac{(\bar{f_1}\bar{f_2}-f_1f_2) V_2(12)+
\mbox{perm.}}{z- \gve_1-\gve_2-\gve_3}\;G_3(z),
\label{eqn:G3}
\end{eqnarray}
where the corresponding single particle self-energy in Hartree-Fock
approximation is given by
\begin{equation}
\gve_1 = \frac{k^2_1}{2m_1} + \Sigma^{HF}(1),
\qquad\Sigma^{HF}(1) = \sum_{2}
\left[ V_2(12,12)-  V_2(12,21) \right]\,  f_2,
\label{eqn:selfHF}
\end{equation}

\begin{figure}[tb]
\begin{minipage}{0.48\textwidth}
\epsfig{figure=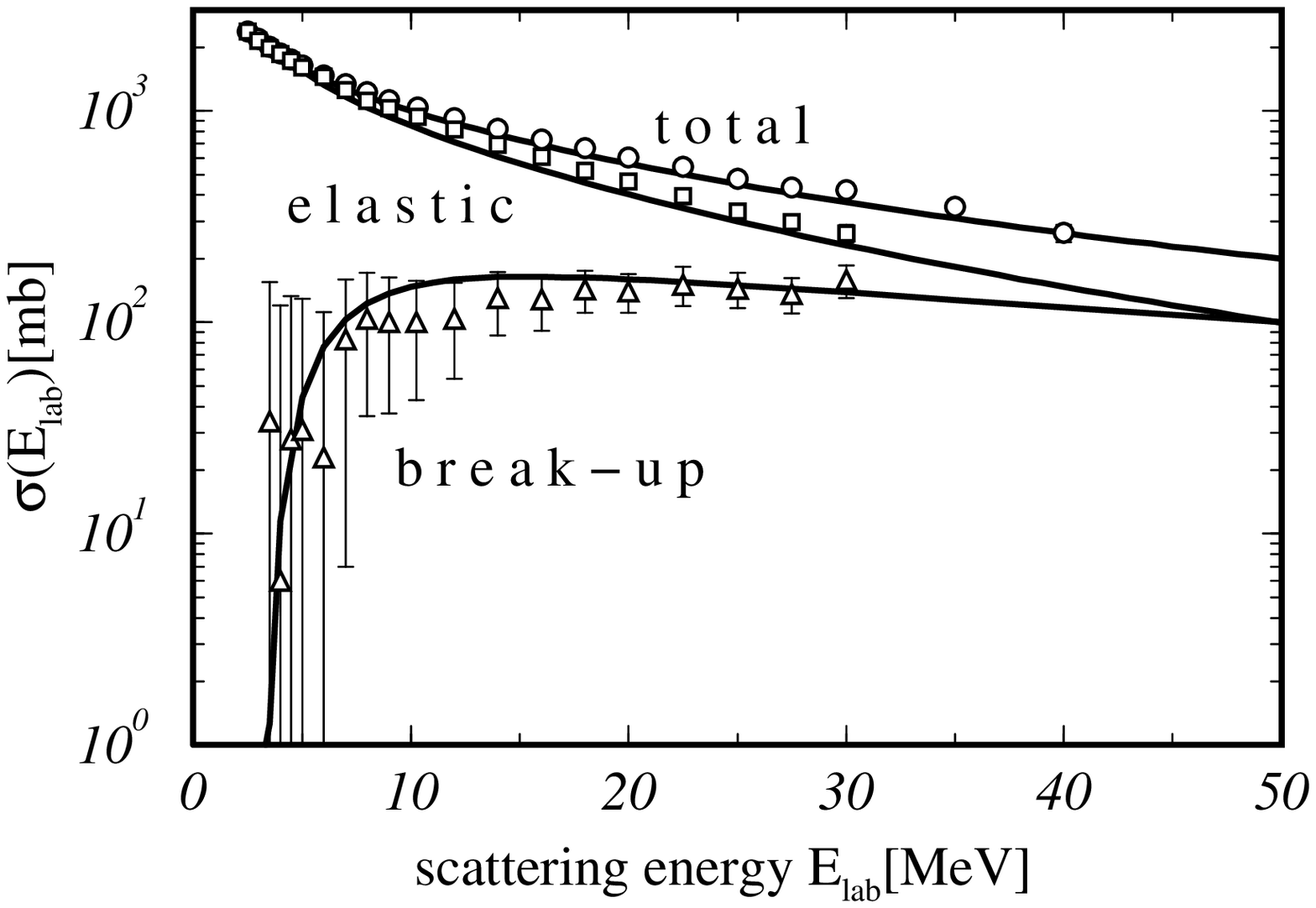,width=\textwidth}
\caption{\label{fig:isocross} Neutron deuteron cross
  sections. Experimental data Schwarz et al.~\protect\cite{sch83}}
\end{minipage}
\hfill
\begin{minipage}{0.48\textwidth}
\epsfig{figure=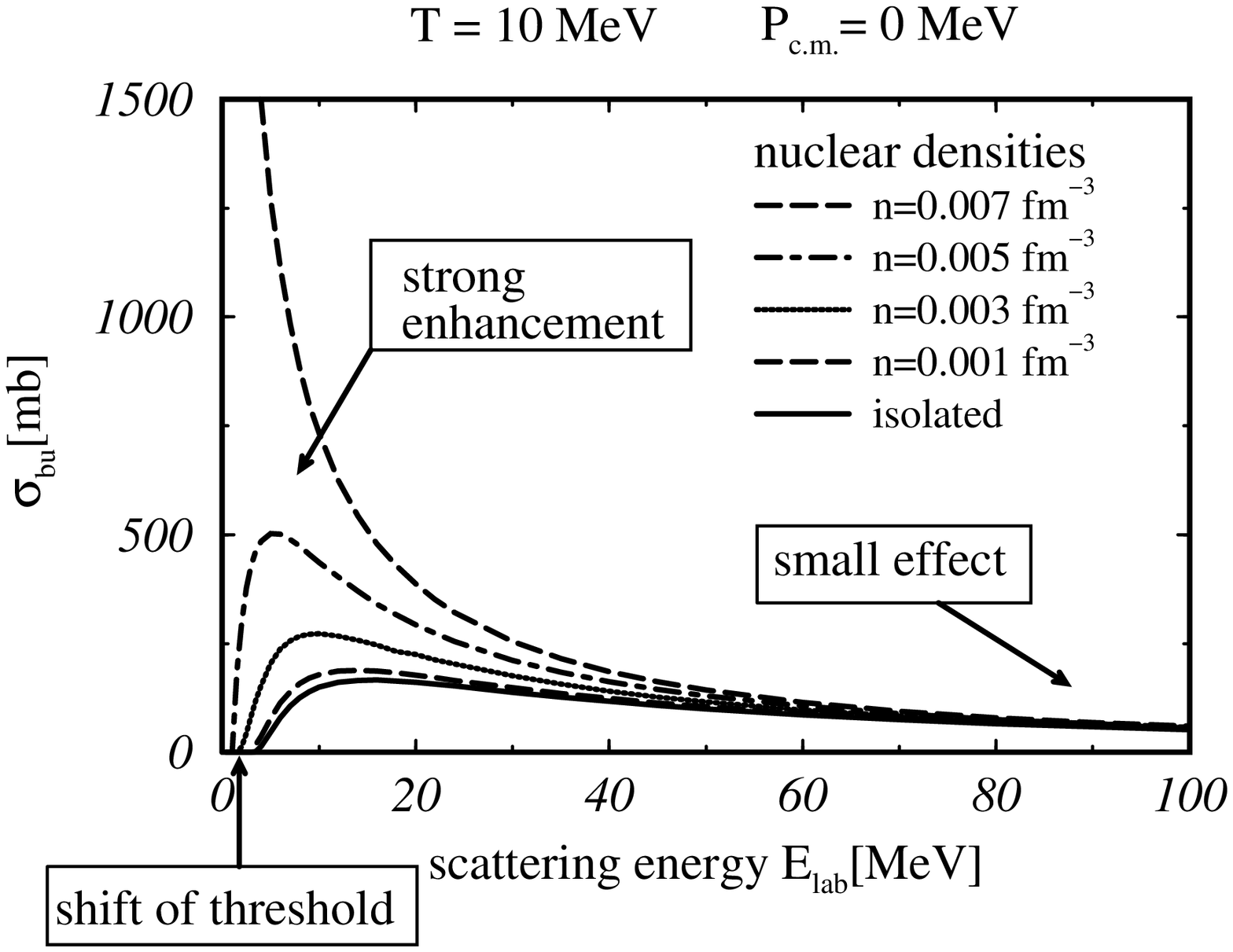,width=\textwidth}
\caption{\label{fig:breakup} Break-up cross section for different
  densities at a  temperature $T=10$ MeV.}
\end{minipage}
\end{figure}
We assume the dominant two-body interaction $V_2$ only. For a recent
and throughout investigation of nuclear three-body forces see
Ref.~\cite{hueber}.  A similar equation has been derived for the
channel\footnote{The three-body fragmentation channels are labeled by
  the spectator particle $\alpha\in\{1,2,3\}$ and the break-up channel
  by zero.} Green function $G_3^{(\alpha)}(z)$, for the two-body
subchannel~\cite{habil}. The AGS transition operator
$U_{\alpha\beta}(z)$ is defined by
\begin{equation}
G_3(z)=\delta_{\alpha \beta}G^{(\alpha)}_3(z)+
G^{(\alpha)}_3(z){ U_{\alpha\beta}(z)} G^{(\beta)}_3(z).
\label{eqn:Galpha}
\end{equation}
For $f\rightarrow 0$ this operator leads to the isolated AGS-operator
with the correct reduction formulas. 
The resulting in-medium AGS equation for $U_{\alpha\beta}(z)$ is then
\begin{eqnarray}\nonumber
 U_{\alpha\beta}(z)&=& (1-{\delta}_{\alpha\beta})
\left[N_3 R^{(0)}_3\right]^{-1}\\[1ex]
&&+ 
\sum_{\gamma\neq \alpha}
N_3^{-1} N^{(\gc)}_2
T_3^{(\gamma)}(z) R^{(0)}_3 N_3\;
U_{\gamma\beta}(z).
\label{eqn:AGS}
\end{eqnarray}
To simplify notation we have introduced 
$N_3=\bar{f_1}\bar{f_2}\bar{f_3}+f_1f_2f_3$ and for
the two-body subchannel of the three-body system
$N_2^{(3)}=\bar{f_1}\bar{f_2}-f_1f_2$ (cycl. perm.) as well as the
three-body resolvent $R^{(0)}_3(z)= (z-{ \gve_1-\gve_2-\gve_3})^{-1}$.
The respective two-body $t$ matrix in the subchannel $\gamma$ entering in
the above equation is
\begin{eqnarray}
T^{(\gamma)}_3(z) &=&   V_2^{(\gamma)} + 
 V_2^{(\gamma)} R^{(0)}_3(z) N_2^{(\gamma)}  T_3^{(\gamma)}(z).
\label{eqn:T2}
\end{eqnarray}
The optical theorem has been used to calculate the break-up, the total
and the elastic cross sections. For the isolated three-body case the
respective theoretical results are shown in Fig.~\ref{fig:isocross}
and compared to the latest experimental data on neutron deuteron
scattering. The calculation is done using a rank one Yamaguchi
potential for the $^1{\rm S}_0$ and coupled $^3{\rm S}_1- ^3{\rm D}_1$
channels. The effect of the nuclear medium is seen in
Fig.~\ref{fig:breakup}. The specific effects on the cross section are
indicated in the figure. Note that the threshold shifts to smaller
energies as it should, since the deuteron binding energy becomes
smaller with increasing density. As the binding energy of the two-body
subsystem approaches zero one might expect an infinite number of
states in the corresponding three-body bound state (Efimov effect).
Note, however, that in the three-body system the binding energy of the
subsystem is not a fixed quantity but depends on the c.m. momentum of
the subsystem, i.e. the binding energy varies in the three-body bound
state and therefore new bound states are not expected to appear.

\section{Reaction time scales}
\begin{figure}[tb]
\begin{minipage}{0.48\textwidth}
\epsfig{figure=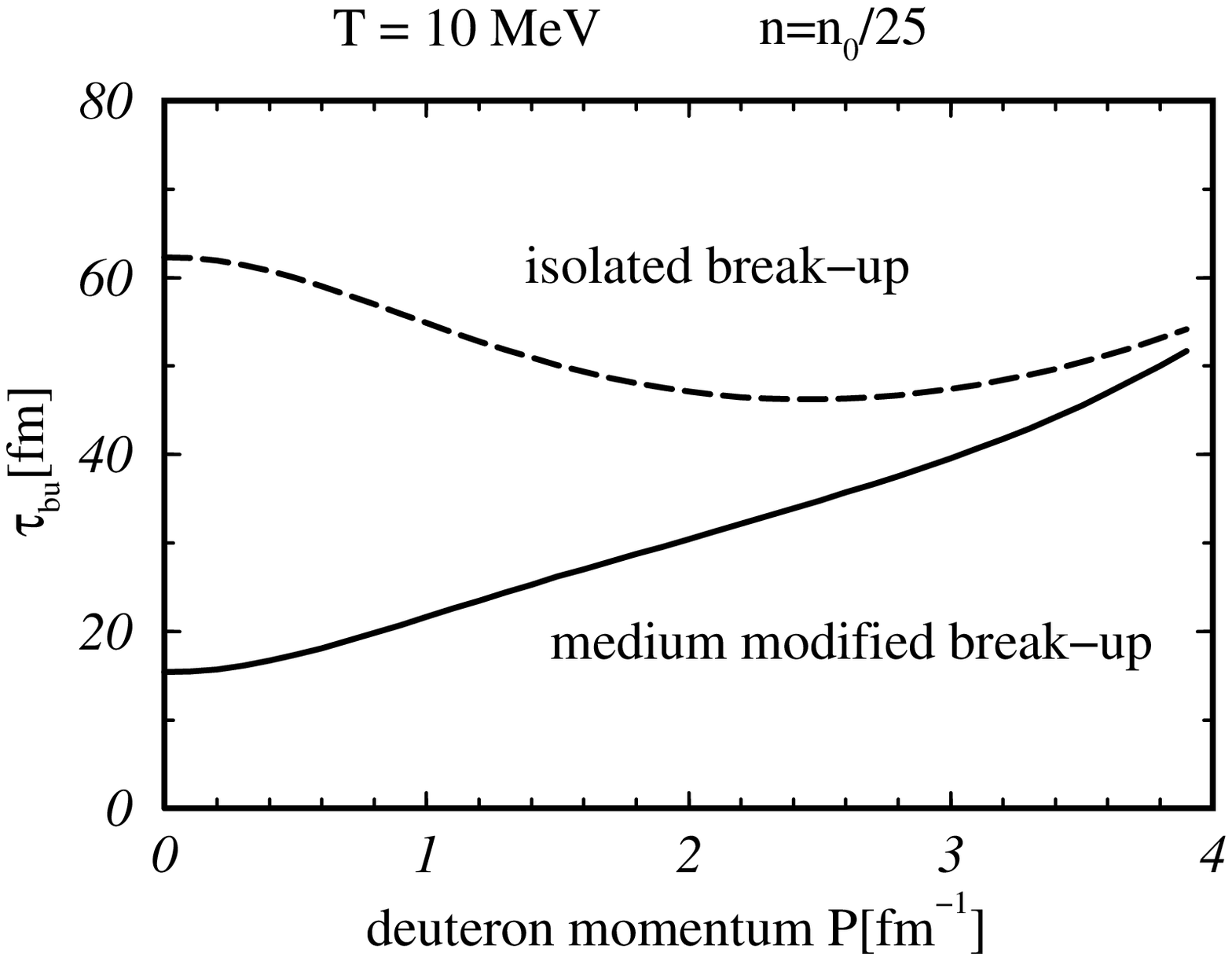,width=\textwidth}
\caption{\label{fig:life} Deuteron break-up time as a function of the
  deuteron momentum.}
\end{minipage}
\hfill
\begin{minipage}{0.48\textwidth}
\epsfig{figure=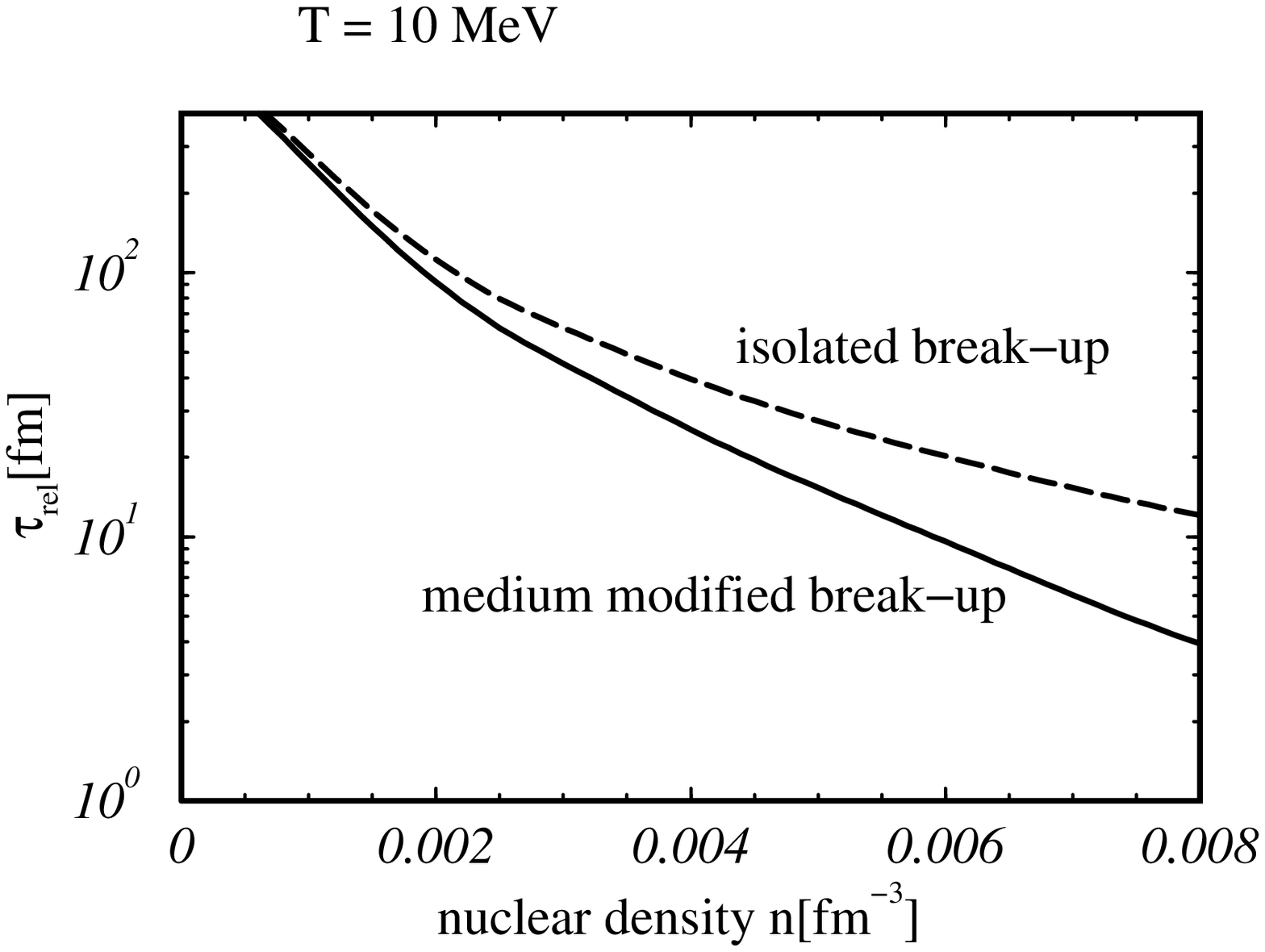,width=\textwidth}
\caption{\label{fig:chem} Chemical relaxation time at $T=10$ MeV as a
  function of nuclear density.}
\end{minipage}
\end{figure}

One consequence of this strongly enhanced cross section can be
directly noticed in the time scales involved. To this end we linearize 
the Boltzmann equation given in (\ref{eqn:Boltz}) as e.g.  done in
Ref.~\cite{bey97} in the context of the Green function
approach. Linearizing with respect to small
variations of the deuteron distribution $\gd f=f^0-f$ from the
equilibrium one $f^0$ leads to
\begin{equation}
\frac{\partial}{\partial \,t} \;\gd f_d(P,t) 
= -\frac{1}{\tau_{\rm bu}[P,n,T]}\;
\gd f_d(P,t),
\end{equation}
where we have introduced the deuteron break-up time $\tau_{\rm
  bu}[P,n,T]$ that depends on the density $n$, temperature $T$, and
deuteron momentum $P$. Using the cross section and the relative
velocities the deuteron break-up time may be written as
\begin{equation}
\tau^{-1}_{\rm bu} = \frac{4}{(2\pi)^3}
\;\int d^3k_N\; |{\mbf{v}}_d-{\mbf{v}}_N|\;{\sigma_{\rm bu}(P)}
\;f_N^0(\gve).
\label{eqn:life}
\end{equation}
The numerical result is shown in Fig.~\ref{fig:life}, where we compare the
use of the isolated to the in-medium break-up cross section in
Eq.~(\ref{eqn:life}) at $T=10$ MeV and a nuclear density of $n=0.007$
fm$^{-3}\simeq n_0/25$, $n_0$ normal nuclear matter density.  The
medium dependent elementary cross section leads to shorter fluctuation
times. As expected the difference is vanishing for larger momenta. 

A similar result is found for the relaxation time to chemical
equilibrium as explained in the following. Linearizing the
corresponding rate equations
\begin{eqnarray*}
\frac{d}{dt}n_d(t)&=&-\alpha (t) n_N(t) n_d(t)+\beta (t) n_N^3(t)\\
\frac{d}{dt}n_N(t)&=&2\alpha (t) n_N(t) n_d(t)-2\beta (t) n_N^3(t)
\end{eqnarray*}
%where the rate coefficient for disintegration $\alpha$ is given by
%$$
%\alpha(t)=
%\int d^3Pd^3k\;
%|\mbf{v}_N-\mbf{v}_d|{\sigma_{bu}}
%\frac{f_N(k,t)}{n_N(t)}
%\frac{f_d(P,t)}{n_d(t)}\nonumber
%$$
with respect to small distortions $\gd n=n^{0}-n$ from the equilibrium
distribution $n^{0}$, viz.
\begin{equation}
\frac{d}{dt}\, \gd n_d(t) = -\frac{1}{\tau_{\rm rel}[n,T]}\,
\gd n_d(t),
\end{equation}
  leads to the corresponding relaxation time
(detailed balance provided)
\begin{equation}
\tau^{-1}_{\rm rel} = \int d^3Pd^3k\;
|\mbf{v}_N-\mbf{v}_d|{\sigma_{\rm bu}(P)}
\frac{f_N(k,t)}{n_N(t)}
\frac{f_d(P,t)}{n_d(t)}\nonumber
\;\left[n_N+4n_d\right].
\end{equation}
In Fig.~\ref{fig:chem} we show the comparison of using the different
cross sections under discussion. Again the difference in time scales
is significant, increasing with increasing density~\cite{kuhrts}.

\begin{figure}[tb]
\begin{minipage}{0.48\textwidth}
\epsfig{figure=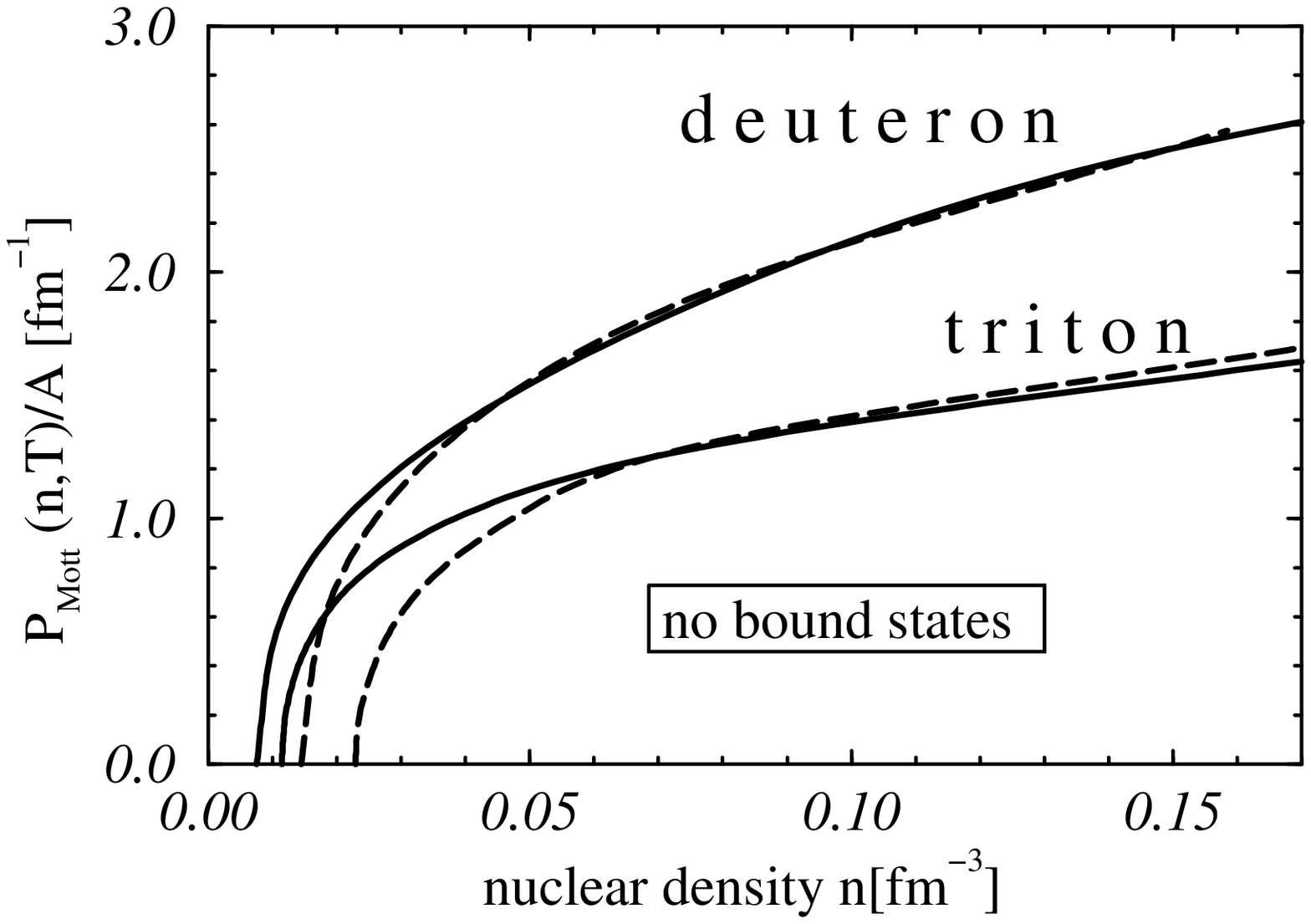,width=\textwidth}
\caption{\label{fig:mott} Mott momentum for deuteron and tritons at
  $T=10$ MeV (solid lines) and $T=20$ MeV (dashed lines).}
\end{minipage}
\hfill
\begin{minipage}{0.48\textwidth}
\epsfig{figure=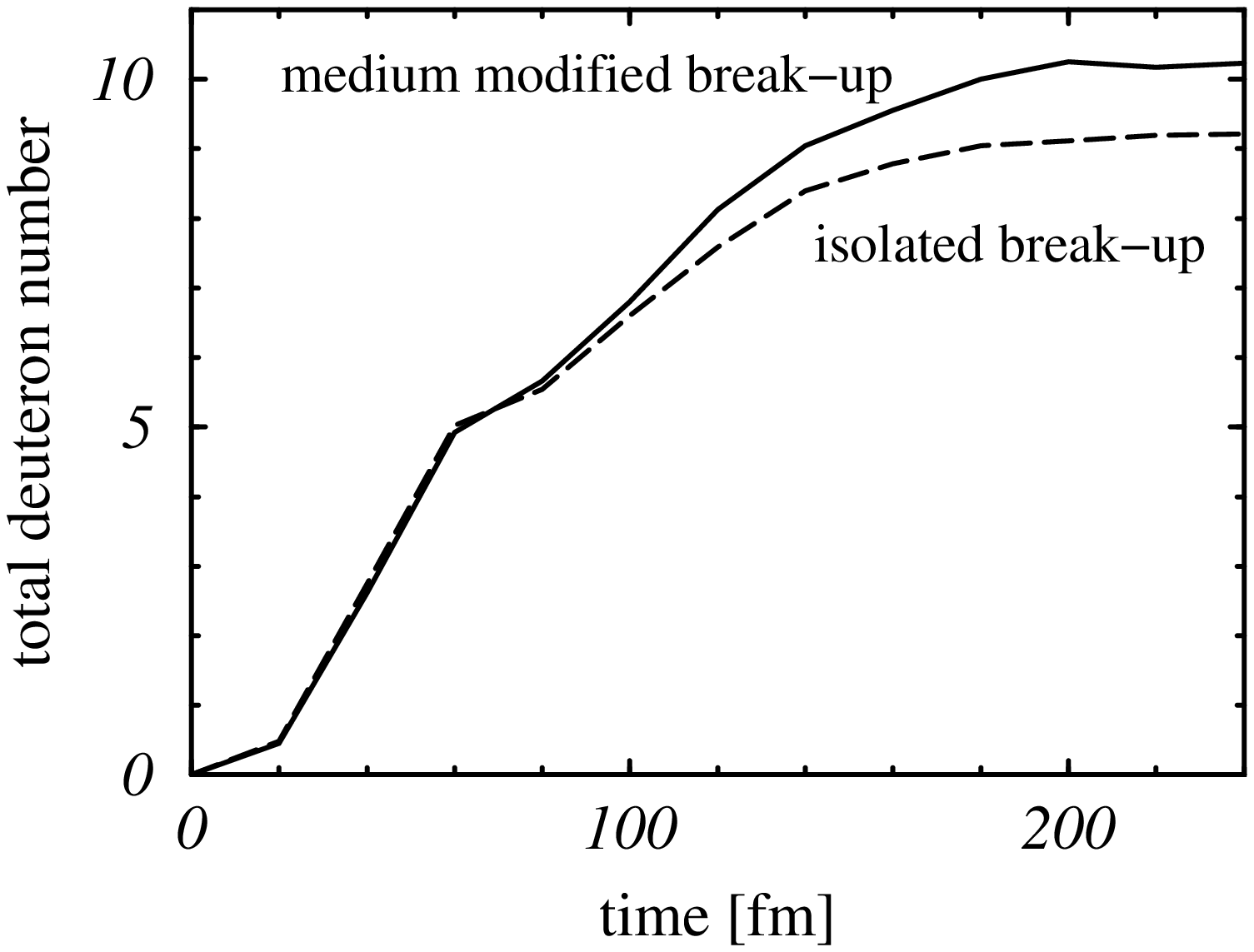,width=\textwidth}
\caption{\label{fig:hic} Total number of deuterons in time. Due to
  statistics and temperature the lines are considered preliminary. }
\end{minipage}
\end{figure}

Another well known many-body feature of bound states is the Mott
effect. For fermions bound states should vanish beyond a certain
density. This density is momentum dependent (Mott momentum) that is
shown in Fig.~\ref{fig:mott}. Below the lines shown no pole exist in
the respective few-body equations. Using the present approach this is
elaborated in detail for the three-body system in Ref.~\cite{schadow}.
The Mott effect is an important ingredient in microscopic simulations
of heavy ion collisions, see e.g.~\cite{dan91}, and may have consequences
for the triton to helium-3 ratio in asymmetric nuclear 
matter.

\section{Extension to heavy ion collisions}
So far the investigation has been restricted to nuclear matter
(homogeneous and symmetric). A laboratory situation closest to nuclear 
matter is provided by heavy ion collisions. As a test case we
investigate the collision $^{129}{\rm Xe} + ^{119}{\rm Sn}$  at 50
MeV/A lab. energy. This collision has been investigated by the INDRA
collaboration, see e.g. Ref.~\cite{indra}.

The density profile suggests that the medium effects influence the
deuteron formation in the final stage of the heavy ion collision,
because of the Mott effect. The temperature in this final stage
extracted from the BUU simulation assuming a Fermi distribution is
rather homogeneous and $T\simeq 4\dots 6$ MeV. This is compatible with
results we achieved with the Quantum Molecular Dynamics code provided
by Aichelin~\cite{aichelin} and other calculations~\cite{wolter}.
Fig.~\ref{fig:hic} refers to a preliminary calculation using the cross
sections shown in Fig.~\ref{fig:breakup} that have been implemented in
the BUU code~\cite{dan91}. The number of deuterons is increasing up to
10\%.  Also, note that the in-medium effect becomes visible at the
final stage only. The density is small enough for
deuterons to significantly survive.

\begin{figure}[tb]
\begin{minipage}{0.48\textwidth}
\epsfig{figure=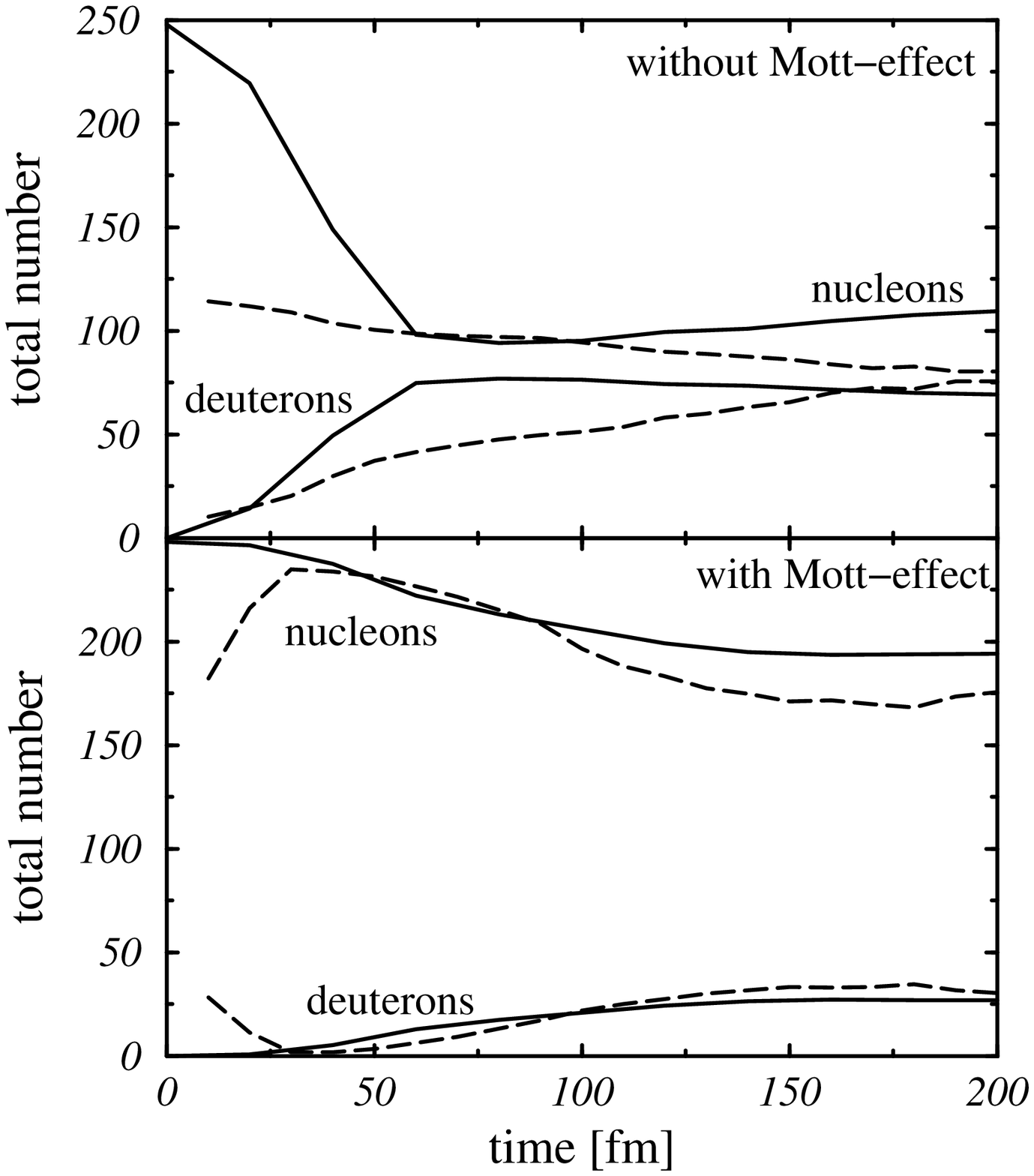,width=\textwidth}
\caption{\label{fig:equi0} Number of nucleons and deuterons in the BUU 
  simulation (solid lines) compared to the respective law of mass
  action (dashed lines).}
\end{minipage}
\hfill
\begin{minipage}{0.48\textwidth}
\epsfig{figure=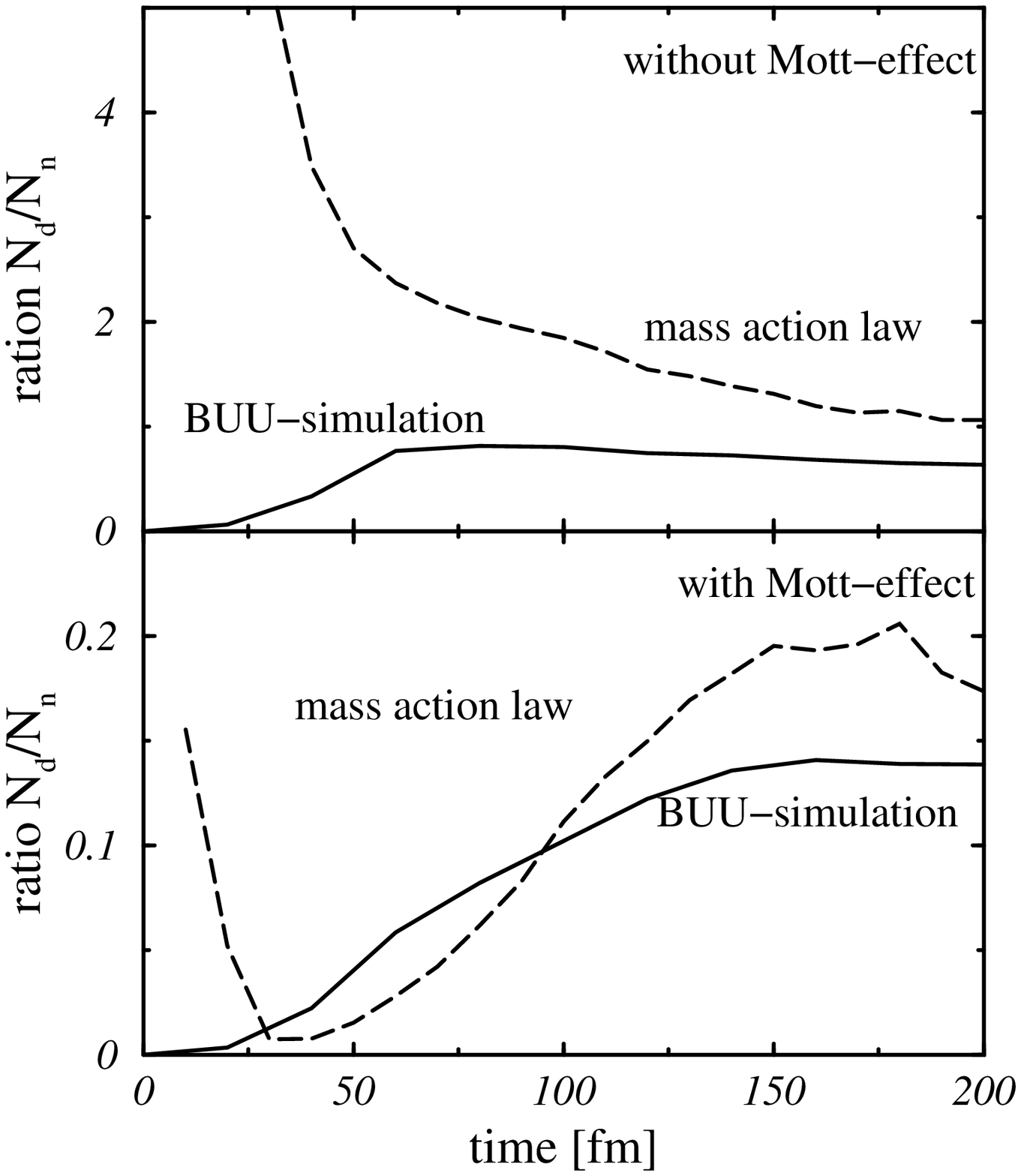,width=\textwidth}
\caption{\label{fig:equi1} Deuteron to nucleon ratio for the BUU
  simulation (solid line) compared to the law of mass action (dashed
  line).}
\end{minipage}
\end{figure}

Finally, we address the question of chemical equilibrium during the
heavy ion collision. Fig.~\ref{fig:equi0} shows the total numbers of
deuterons and nucleons as given in the BUU simulation (solid line)
compared to the law of mass action result. The upper figure ignores
the Mott effect (that would lead to a wrong description of the
experimental data), whereas the lower figure shows the same with the
Mott effect properly taken into account. The calculation is done
without medium modifications for simplicity. Including the Mott effect
reduces the number of final deuterons by about a factor of 3 and in
turn increases the number of nucleons by about a factor of two. In the
deuteron to nucleon ratio this ends up to about a factor of 6 shown in
Fig.~\ref{fig:equi1} for the same process. It seems that for the
process considered ($^{129}{\rm Xe} + ^{119}{\rm Sn}$ at 50 MeV/A lab.
energy central collision) the evolution is rather close to chemical
equilibrium provided the Mott effects are properly accounted for.

\section{Conclusion and Outlook}
In summary, we have presented a systematic and consistent decoupling
scheme to handle the many-body problem. This leads in turn to
in-medium few-body equations that have been rigorously solved for the
three-body system to treat the deuteron formation via three-particle
collisions. The prominent changes induced by the medium are the {\em
  self-energy shifts}, the {\em Pauli blocking}, the {\em Mott effect}
and the change of {\em reaction rates}. Within linear response we have
shown that the change in the rates leads also to {\em faster time
  scales}.  The deuteron production for the BUU simulation of the
$^{129}{\rm Xe} + ^{119}{\rm Sn}$ at 50 MeV/A lab. energy central
heavy ion collision is {\em enhanced by 10\%}.

The exact treatment of few-body systems embedded in the medium is
however not restricted to nuclear physics. Further potential
applications are possible, e.g., in the field of semiconductors to
treat the formation of excitons or the trion bound state in the dense
low dimensional electron plasmas. For recent experiments see
e.g.~\cite{trion}. For highly ionized dense plasmas the impulse
approximation may fail for the three-particle break-up cross section
needed to calculate the ionisation and recombination rates.

Finally, the formalism is capable to also treat the four-particle
problem. In this context the $\alpha$-particle is the most interesting
object.  Because it is light and has a large binding energy per
nucleon, it should play a dominant role in the decompressing hot
nuclear matter.

{\em Acknowledgments.} We gratefully acknowledge many discussions on
the topic with P. Schuck (Grenoble), a fruitful collaboration with W.
Schadow (TRIUMF) on the triton, and A. Schnell (UWA Seattle) for
providing us with a code to calculate the nucleon self energy.

\noindent\rule{\textwidth}{0.2pt}

\end{document}